\documentclass[11pt,a4paper]{article}

\usepackage[latin1] {inputenc}
\usepackage{amsmath}
\usepackage{amsfonts}
\usepackage{amssymb}

\usepackage[latin1] {inputenc}
\usepackage{amsmath}
\usepackage{amsfonts}
\usepackage{amssymb}
\usepackage[dvips]{graphicx}
\usepackage{psfrag}
\usepackage{multicol}

\newtheorem{definition}{Definition}

\newtheorem{lemma}[definition]{Lemma}
\newtheorem{fact}[definition]{Fact}

\newtheorem{theorem}[definition]{Theorem}
\newtheorem{cor}[definition]{Corollary}

\newcommand{\conf}[1]{\mbox{{\sc conf}} \left( #1 \right)}
\newcommand{\Red}[1]{\mbox{ \textrm{Red}} \left( #1 \right)}
\newcommand{\Redc}[1]{\mbox{\textrm{Red-close}} \left( #1 \right)}
\newcommand{\Prob}[1]{\mathbf{P} \left( #1 \right)}
\newcommand{\Expec}[1]{\mathbf{E} \left[ #1 \right]}
\newcommand{\sQ}{\mathcal{Q}}

\newcommand{\cD}{D_{\mbox{{\tiny $\Box$}} }}
\newcommand{\cd}{d_{\mbox{{\tiny $\Box$}} }}
\newcommand{\sB}{\mathcal{B}}
\newcommand{\sG}{\mathcal{G}}
\newcommand{\sC}{\mathcal{C}}
\newcommand{\sS}{\mathcal{S}}
\newcommand{\cc}{\mbox{\sc{c}}}
\newcommand{\cw}{\mbox{\sc{w}}}
\newcommand{\cz}{\mbox{\sc{z}}}

\newcommand{\area}{\mbox{\sc{area}}}
\newcommand{\ecc}{\mbox{\sc{ecc}}}

\newcommand{\nR}[2]{\#_ r^{#2}( #1 )}
\newcommand{\nW}[2]{\#_w^{#2}( #1 )}
\newcommand{\nB}[2]{\#_b^{#2}( #1 )}

\newcommand{\bb}{\mbox{\sc{b}}}
\newcommand{\ww}{\mbox{\sc{w}}}
\newcommand{\rr}{\mbox{\sc{r}}}
\newcommand{\poly}{\mbox{\sc{poly}}}

\newcommand{\qed}{\hspace{\stretch{1}$\square$}}

\newcommand{\ideaproof}{\noindent\textit{Idea of the proof. }}

\begin{document}
\title{\textbf{  Parsimonious Flooding   in     Geometric Random-Walks }\thanks{This work was partially supported by the PRIN 2008 research project COGENT (COmputational and GamE-theoretic aspects of uncoordinated NeTworks), funded by the Italian Ministry of Education, University, and Research.}}

\author{
Andrea  Clementi \\
{\small Dipartimento di Matematica} \\ {\small  Universit\`a   ``Tor Vergata'' di Roma}  \\ {\small  clementi@mat.uniroma2.it} 
\and Riccardo Silvestri \\  {\small Dipartimento di Informatica} \\ {\small  Sapienza Universit{\`a}  di Roma } \\
{\small silvestri@di.uniroma1.it }
}

\maketitle

\begin{abstract} 
We study the information spreading yielded by the \emph{(Parsimonious) 
$1$-Flooding Protocol} in  geometric Mobile Ad-Hoc Networks. 
We consider $n$ agents on    a convex 
plane region of diameter $D$  performing independent random walks with  move radius $\rho$. 
At any time step, every active agent $v$ informs 
every non-informed agent  which is within distance $R$ from $v$ ($R>0$ is the 
transmission  radius). An agent is only active at the time step immediately after the 
one in which 
has been informed and, after that, she  is removed. 
At the initial time step, a source agent is informed and we look at the 
\emph{completion  time} of the protocol, i.e., the first time step (if any) in which all 
agents are  informed. 

\noindent 
This random process is equivalent to the well-known 
\emph{Susceptible-Infective-Removed  ($SIR$}) infection process in 
Mathematical Epidemiology. No analytical results are available for this 
random  process  over any  explicit  mobility model. 

\noindent 
The presence of removed agents makes this process 
much more complex than the (standard) flooding. 
We prove optimal bounds on the completion time depending on the parameters 
$n$, 
$D$, $R$, and $\rho$. The obtained bounds hold with high probability. 

\noindent 
We remark that our method of analysis provides a clear picture 
of the dynamic shape of the information spreading (or infection wave) 
over the time. 

\end{abstract}

\medskip

\thispagestyle{empty}
\newpage

 \section{Introduction}

In the \emph{Geometric Random-Walk Model}  \cite{CMPS09,DMP08,GT02},   $n$ agents perform
   independent random walks
  over a bounded  plane  region $\sS$ and  we consider  the following  \emph{infection} process.
 Every agent can be in three different states:  non-informed (white) state, informed-active (red),
informed-removed (black).  During a time step, every agent performs one step of the random walk and 
every red agent informs (infects)  all  white agents lying within distance $R$. A white agent that has been informed (for the first time) at time step $t$, she   becomes
red at time step $t+1$.
Finally, when  an agent becomes red   she   stays red    for just one  time step and, after that, she    becomes  black  
 and stays that,    forever.

\noindent
 At the initial time step, a source agent is in the red  state. The completion time of the above infection process is  the first time step in which every agent gets into the black state. If this time step does not exist then we say 
   the infection process does not complete.

\noindent
This   random process  is inspired by   two main scenarios:  The   \emph{Susceptible-Infected-Removed  (SIR)} process which is 
widely studied in  Mathematical Epidemiology \cite{BN02,BN08,BDW08} and the 
    \emph{(Parsimonious) $1$-Flooding Protocol} \cite{BCF09} in  \emph{geometric Mobile Ad-Hoc Networks (MANET)}.
    The    (parsimonious) $k$-flooding protocol is the generalized version of the above infection process where a white  agent, which is informed at time $t$, will remain  red (so active) for the next $k$ time steps. 
 
  While the standard flooding is extremely inefficient in terms of agent's
energy consumption and message complexity,
the    $k$-flooding protocol (for small values of $k$)   strongly  reduces    the agent's energy consumption
and   the overall message complexity.
However, as discussed later,
the infection process yielded by the   $k$-flooding (for small $k$) over a dynamic network is much more complex than that
yielded 
  by the standard flooding.  

 The   $k$-flooding   has been  
  studied in \cite{BCF09} on \emph{ Edge-Markovian Evolving Graphs} (\emph{Edge-MEG}).  
    An Edge-MEG \cite{CMMPS08,CMPS09} is a Markovian 
  random process that generates an infinite sequence of graphs over the same set of     $n$ agents.
   If an edge exists at time $t$ then, at time $t+1$, 
it dies with probability $q$. If instead the edge does not exist at time $t$, then it will come into existence at time $t+1$ with probability $p$.
 The stationary distribution of an Edge-MEG with parameter $p$ and $q$  is the famous Erd\"os-R\'enyi random graph $\sG(n,\tilde p)$ where $\tilde p = \frac{p}{p+q}$.
The work \cite{BCF09} gives   tight bounds on the   $k$-flooding  on stationary Edge-MEG  for arbitrary values of $p$, $q$ and $k$.
In particular, it derives the \emph{reachability threshold} for the  $k$-flooding, i.e., the smallest $k=k(n,p,q)$ over which the   protocol completes.
  
 \noindent
  Edge-MEG  is an   analytical model of dynamic networks capturing time-dependencies, an important feature observed in real scenarios such as  (faulty)  wireless networks and P2P networks. However, it does not model important features
  of  MANET.  Indeed, in Edge-MEG,
   edges are independent  Markov chains while, in MANET, there is a strong correlation among edges:   agents
   use to   act over a geometric space \cite{CMPS09,DMP08,GT02,LV05}.

\medskip
\noindent
\textbf{Our Contribution. }
We  study the   $1$-flooding protocol in    the geometric random-walk model
(the resulting MANET   will be   called  \emph{geometric-MANET}).
The move radius $\rho$ determines the maximal distance an agent can travel
in one time step.
 Even though   network modelling  issues  are out of the
aims of this work, both the transmission radius and the move radius play a crucial role in our analysis and, thus, a small discussion
about them is needed.
Both parameters  depend on several factors. In mathematical epidemiology, they  depend on the agent's mobility, 
 the kind of infection and  on  the   agent's social behaviour    that all together 
 determine the average rate   of ``positive'' contacts.
 In  typical biological cases, the move radius can be significantly larger than the transmission radius.
 In MANET,  the move radius   depends on, besides  the agent's mobility, 
     the   adopted protocol that tunes the  transmission rate of the agents:
 the larger is the time between two consecutive transmissions the larger is $\rho$.
A  larger $\rho$ could  yield  a better  message complexity   at the cost of
a larger completion time  (the correct trade-offs is derived      in \cite{CPS09} for the standard flooding).  
  
  \noindent
  It turns out that setting the   ``most convenient'' value
of the move radius is an     issue that concerns both the network modelling and the protocol design.
For these reasons,   we investigate the   $1$-flooding for a wide  range of  values for parameters  
$R$ and $\rho$.

\noindent
Most of the known  analytical studies concern   geometric-MANET defined  over    squares or disks.
We instead consider any   \emph{measurable} region $\sS$ of diameter $D(\sS)$.  Informally speaking,
 a bounded region of the plane is said \emph{measurable}
if it is convex and it can be well approximated by a cover of square cells of  suitable  size. Furthermore, 
in a measurable region,   we require 
   the       geometric random-walk model   yields an almost-uniform stationary distribution of the agents.  
Square and disks are in fact the simplest examples of  regions having all such properties.

\noindent
We first show a negative result. Given a measurable region $\sS$, if the transmission radius $R$ is below 
 the
\emph{connectivity threshold}\footnote{The connectivity threshold refers to
the uniform   distribution of $n$ (static) agents over $\sS$. } of $\sS$,
 then, for any    $\rho \geqslant 0$, w.h.p.\footnote{As usual, an event is said to hold \emph{with high probability (w.h.p.)}
 if its probability is at least    $1-1/n$. }  there are $m = n^{\alpha}$ (for some   constant  $\alpha >0$)
   possible     source agents for which the 1-flooding does not complete. 
    This negative result also holds for the $k$-flooding for any
$k = O(1)$ and $\rho \leqslant R$. 

\noindent
We thus study the   1-flooding for $R$ over  the connectivity threshold of $\sS$.
We emphasize that, evenin this case, nothing is known  about
the \emph{connectivity }   of the subset of red agents over the time and, more generally, 
 about the impact of agent mobility on the 1-flooding process.
 
 \noindent
If  $\rho \leq R/(2\sqrt 2)$, we prove that
the information spreads at ``optimal'' speed $\Theta(R)$, i.e., the  1-flooding protocol w.h.p. 
 completes  within $O(D(\sS)/R)$ time steps.
Observe that, since $\rho < R$, this bound is   asymptotically optimal. 
 
\noindent
Then, we consider  move radii that can be up to any polynomial of    $R$, i.e.  $ \rho \leqslant \poly(R)$.
We prove that the information spreads at ``optimal'' speed   that is  $\Theta(\rho)$.
So the 1-flooding w.h.p. completes in time $O(D(\sS)/\rho)$ which is optimal for any $\rho \geqslant R$.
Notice that  this  optimal information speed  makes the 1-flooding time   smaller than the static diameter
$O(D(\sS)/R)$ of the stationary graph:   our bound 
is thus  the first  analytical evidence that  agent's mobility   actually speeds-up this infection process.  
Finally, we observe that, in both cases,  the energy-efficient
1-flooding  protocol  is as   fast as the  standard   flooding     \cite{CMPS09,CPS09}.  

\noindent
Can any   upper bound for the 1-flooding time  be directly extended to the $k$-flooding time, for any $k>1$? 
Rather surprisingly, the answer is in general 
 not positive. Indeed, consider the two protocols running independently on the same experiment of the  geometric-MANET.
  In the $k$-flooding process, an agent might be informed much earlier than what happens
in the 1-flooding one: hence   she might turn into the black state   earlier  and this might be fatal for some other
white agents meet later. So a    coupling between
the  two processes  does not seem to exist. Another surprising fact is that, for the same reasons,  increasing the transmission
 $R$ does not necessarily imply the same or a better completion time of the $k$-flooding, for any $k$.  In other words, 
 it is not easy  to establish whether 
the $k$-flooding  time  is  a  ``monotone'' function with respect to  $k$ or $R$. 
  These facts provide a good insight  about the crucial   differences between
the flooding  process  and the $k$-flooding  one. Such differences make the    analysis techniques  adopted 
for   the flooding    almost useless. In particular,
 percolation theory   \cite{GT02,DMP08,PSSS10}, meeting   and cover time of random walks on  graphs  
  \cite{PPPU10},
and the expansion/bootstrap arguments   \cite{CMPS09,CPS09} strongly rely 
on the fact that an informed agent will be active for  all the flooding  process.   
Furthermore, the analysis of $k$-flooding over the Edge-MEG model  \cite{BCF09} strongly relies on the stochastic  independence  
among the   edges in  $\sG(n,p)$: a   property that clearly does not hold in geometric-MANET. 

\noindent
Our method of analysis   significantly departs from all  those mentioned above.
Besides the optimal bounds on the completion time, our analyses   provide a clear characterization of the geometric
evolution of the infection process. We   make use of  a   grid partition of $\sS$ into   cells of size $\Theta(R)$
and  define a set of possible states a cell can assume over time 
 depending  on the number of red, white and black agents inside it.  We then derive  the local state-evolution law of any cell.
 Thanks to the regularity of   this  law, 
  we can    characterize   
    the   evolution of the  geometric wave formed by the \emph{red cells}  (i.e. cells
containing some red agent). A crucial   property we prove  is that, at any time step,  white cells (i.e. cells containing  white agents only) will never be adjacent to black cells,
 so there is always a red  wave working  between the black region and the white one.
Furthermore, we show that  the red wave eventually spans     the entire  region before all agents become black.

\noindent
The generality of our method of analysis has   further   consequences.
Thanks to the  regularity of the red-wave  shape, we are 
 able to bound the time by which a given subregion   will be infected   for the first time. This bound is 
  a function of the distance between the subregion  and the initial position of the source agent. Actually, it is 
  a function of the distance from the \emph{closest}  red agent at the starting time.  So, our technique also works in the 
   presence
  of an arbitrary set    of source agents that aim to spread the same infection (or message).
  Under the same assumptions made for the single-source case, we can prove   
  the completion time is    w.h.p.   $\Theta(\ecc(A,\sS) / R)$ (or $\Theta(\ecc(A,\sS)/\rho)$) where $A$ 
  is the set of the positions of the source agents at starting time and 
 $\ecc(A,\sS)$ is the\emph{ geometric eccentricity} of $A$ in $\sS$.
 Finally,  the specific arguments of our proofs  can be easily adapted to get the same upper bounds
for the $k$-flooding as well.

  \medskip
\noindent
\textbf{Related Works.}
 As mentioned above, there are no analytical study of the parsimonious flooding process  over
 any geometric mobility model.
 In what follows, we briefly discuss    some  analytical results concerning the flooding   over some models  of MANET.
In \cite{KY08},   the flooding  time is studied  over a restricted geometric-MANET. 
Approximately, it corresponds to the case  $\rho = D$. Notice that, 
under this restriction, the  stochastic dependence between two consecutive agent  positions is negligible.
In \cite{JMR09},  the speed of data communication between two agents is studied over a   class of  \emph{Random-Direction} models
yielding uniform  stationary agent's distributions  (including the  geometric-MANET model).
They provide an upper bound on this speed that can be interpreted as a 
  \emph{lower } bound on flooding   time when the mobile  network is very sparse
and disconnected (i.e. $R,\rho = o(1)$). 
Their   technique, based on
Laplacian transform of independent journeys,    cannot be extended to provide  
any upper bound on the time of   any version of  the   flooding.
 We observe that, differently from our model,  both  \cite{KY08} and \cite{JMR09} assume that when an agent
  gets  informed then
 all agents  of her  \emph{current} connected component (no matter how large it is) will get informed in one time step.
 The same unrealistic assumption is adopted in \cite{PSSS10}, where bounds on flooding time (and some other tasks) are obtained   in the   Poisson approximation of the  geometric-MANET model.  
 In \cite{CMPS09,CPS09}, the first almost tight bounds for the flooding time over      geometric-MANET    have been given.
 As mentioned before, their proofs strongly rely on the fact that informed agents stay always   active.
   Flooding and   gossip time for  random walks over the grid graph
  (so,  the agent's space is     a graph) have  been studied in \cite{PPPU10}. 
 Here, an agent  informs all agents lying  in the same node.
 Besides the differences between 1-flooding and flooding discussed before, it is not clear whether    their results could  be
  extended to      the    random walk model over geometric spaces.  Especially, in their model, there is no way to 
 consider arbitrary values of the  move radius.

  \medskip
  \noindent
  \textbf{Roadmap of the paper.}
  The rest of the paper is organized as follows.
  After giving some   preliminary definitions in Section \ref{sec::prely}, we analyze the case $\rho \leqslant R/(2\sqrt 2)$ in 
  Section \ref{ssec::slow}. In Section \ref{sect::high0}, we present the results for 
  the case $\rho \leqslant R^2/\sqrt{\log n}$. In Section \ref{sect::high}, we introduce a slight different
  geometric random-walk  model  and a more powerful technique   to extend
  the bound $O(D(\sS)/\rho)$ to any $\rho \leqslant \poly(R)$. In Section \ref{sec::neg}, we first briefly describe
  the extension to multi-source case and, then, derive  the completion threshold of $R$
         for the 1-flooding. 
     Finally, some    open questions are discussed in Section \ref{sec::conc}. Most of the proofs
     are given in the Appendix.
  
\section{ Preliminaries} \label{sec::prely}
We first introduce the class of   regions that can be considered 
  feasible support spaces for our mobile agents.

\begin{definition}\label{def::measreg}
For any $l,\gamma>0$, a  bounded    connected plane 
region $\sS$ is $( \ell, \gamma)$-measurable if it is convex and 
there exists a   grid-partition of the plane into
squared cells of side length $\ell$ in which  a cell cover $\sC(\sS)$ of $\sS$ exists such that, 
 for any cell $\cc \in \sC(\sS)$,   $\area(\cc \cap \sS) \geqslant \gamma \ell^2$.
\end{definition}
  
  \noindent
  We say that two cells are adjacent if they   touch each other by side or by corner. 
   The  \emph{cell-diameter}
  $\cD(\sS)$  of an $(\ell, \gamma)$-measurable $\sS$  is defined as follows.
   Given two cells $\cc,\cc' \in \sC(\sS)$, define their \emph{cell-distance}
  $\cd(\cc,\cc')$ as the length of the shortest   cell-path  $p = \langle\cc =  \cc_0,\cc_1, \ldots, \cc_s=\cc'\rangle$ such that
 for every $i$,  $\cc_i \in \sC(\sS)$ and $\cc_i$ is adjacent to $\cc_{i+1}$.  Then, we define
 
 \[  \cD(\sS) \ = \ \max\{ \ \cd(\cc,\cc') \ | \ \cc , \cc' \in \sC(\sS)    \  \} \]
 Similarly,  we can define the \emph{cell-distance} between 
a cell and any cell subset $\sC$, i.e., 

\[   \cd(\cc, \sC) \ = \ \min\{  \cd(\cc,\cc ') \ | \ \cc ' \in \sC \} \]

\noindent
 Observe that, since  $\sS$ is convex,   the  (Euclidean)  diameter $D(\sS)$  and the 
 \emph{cell-diameter} $\cD(\sS)$ are tightly related:  $ \cD(\sS)= \Theta(D(\sS) /\ell)$.
 For instance, if $\sS$ is the
 $L\times L$-square ($L>0$) and $\ell  = o(L)$, then its cell diameter is $\Theta(L/\ell)$.
  
  \noindent
  According to the   geometric random-walk model, there are  $n$ agents  that perform    independent random walks
   over    $\sS$.   At any time step, 
an agent in position $x \in \sS$,  can move uniformly at random to any position  
  in   $\sB(x,\rho) \cap \sS$, where $\sB(x,\rho)$ is the disk 
of center    $x$ and    radius $\rho$. 
  This is a special case of the \emph{Random Trip Model} introduced in \cite{LV05} where it is proved that   it admits
 a unique stationary agents distribution.  In the sequel, we always assume that at time $t=0$ agents' positions  are random w.r.t. the stationary distribution.
 
 \noindent
 For the sake of simplicity, we assume that the average density of agents in $\sS$ (i.e. the average number of agents
 per area unit) is 1.  It is straightforward to scale all the definitions and results to an
  arbitrary average  density of agents.

  \noindent
Let us consider $n$ mobile agents acting over an $(\ell,\gamma)$-measurable region $\sS$. 
We say that the resulting geometric-MANET satisfies the  \emph{density condition}    if, for every time step 
$t = 0, 1,\ldots, n$ and  for every cell $\cc \in \sC(\sS)$, the number $\#_{\cc}$ of agents in $\cc$ 
at time step $t$ satisfies, with probability at least $1-(1/n)^4$,
the following inequalities

\begin{equation}\label{eq::dens}
\eta_1 \ell^2  \ \leqslant \ \#_{\cc} \ \leqslant \ \eta_2\ell^2 , \mbox{where $\eta_1,\eta_2$ are positive constants. }
\end{equation}

\noindent
We will consider geometric-MANET that satisfy the density condition. For instance, if the region is the  
$\sqrt n \times \sqrt n$-square, this assumption  implies     $R\geqslant \beta \sqrt{\log n}$,
 for a suitable constant $\beta >0$.
 More generally,
  the  geometric  random-walk model with $n$ agents over an arbitrary  
 $(\ell, \Theta(1))$-measurable region with $\area(\sS)/\ell^2 \leqslant \beta n/\log n$ for a suitable constant $\beta>0$ and
 move radius $\rho = O(\ell)$  satisfies the density condition. 
 The proof of this fact can be obtained by exploiting the same arguments used in \cite{CMPS09,CPS09}
 for  the  discrete approximation of  the  geometric  random-walk model   on the   square.

 According to the 1-flooding protocol, at any time step, every agent can be in three different states:    white (non-informed) state,   red (informed-active),
and  black  (informed-inactive). 
 The configuration $\conf{t}$ is defined
  as the set of the positions of the $n$ agents together with their 
respective states at the end of  time $t$.   
 
\noindent
The  analysis of the information spreading    will be performed
in terms of the number of  infected cells.
Given a cell   $\cc \in \sC(\sS)$, the
neighborhood $N(\cc)$   is the set of cells formed by $\cc$ and all its  adjacent
cells.

   let us observe that, the density condition implies that w.h.p.  the  (geometric) source  eccentricity in $\sS$
    is $\Omega(D(\sS))$; morever  
the maximal speed of any message in the geometric-MANET  is $R+\rho$.
We thus  easily get the following lower bound.

\begin{fact} \label{tc::low}
For any $k$, the $k$-flooding time is w.h.p. $\Omega(D(\sS)/(R + \rho))$.
\end{fact}

\section{  High Transmission-Rate or  Low-Mobility } \label{ssec::slow}

We warm-up with  the
 case where the move radius is smaller than the transmission radius. More precisely,
we assume  $\rho \leqslant R/(2\sqrt 2) $.    Under this assumption,  
an agent lying  in a cell $\cc$ cannot escape from   $N(\cc)$ in one step. 
This makes the analysis   simpler than the 
cases with larger $\rho$. However, some of  the crucial 
ideas  and arguments adopted here will be exploited for the harder cases too.

  \noindent
 We can now state the optimal bound on 1-flooding.
 Consider  a      network formed by   $n$ mobile agents with transmission radius $R$   and move radius 
 $\rho \leq R/(2\sqrt 2)$. The agents act over 
  an $(\ell,\gamma)$-measurable region $\sS$ of   diameter $D(\sS)$ such that  
  $\ell = R/(2\sqrt 2)$, and $\gamma$ is any positive constant. Assume also that the resulting geometric-MANET
  satisfies the density condition.

 \begin{theorem} \label{thm::uppslow}
 Under the above assumptions, the 1-flooding time
  is w.h.p.       $\Theta(D(\sS)/R)$. \end{theorem}

\noindent
The geometric-MANET defined over 
an $L \times L$-square $\sQ$ satisfies the above conditions provided that 
$R\geqslant c_0 L\sqrt{\log n / n}$ for  a  sufficiently large constant   $c_0$ (see \cite{CMPS09,CPS09}).
We thus have the following 

\begin{cor}\label{thm::upp bound}
Let   $R \leqslant L$ be such that $R\geqslant c_0 L\sqrt{\log
n / n}$ for  a  sufficiently large constant   $c_0$ and let  $\rho \leqslant R/(2\sqrt 2)$.  Then the  1-flooding time
over $\sQ$
 is    w.h.p.   $ \Theta (L/ R)$.
\end{cor}

\noindent

  \subsection{Proof of Theorem \ref{thm::uppslow}}

For the sake of simplicity, 
we assume that the time step is divided into 2 consecutive phases: the \emph{transmission phase} where
every red agent transmits the information and the \emph{move phase} where every 
agent  performs one  step of the random walk.
We need to  introduce the feasible states of a cell during the infection process.

\begin{definition} \label{def::cellstate1}
  At (the end of) any time step a cell $\cc$ can be in 4 different states. It is  white  if   it contains only white agents.
It is  red if  it contains at least one red agent. It is black if  it contains black agents only.
It is grey if it is not in any of the previous 3 cases.
\end{definition}

\noindent
We will show that the  infection  process, at any time step,  has (w.h.p.) a well defined   shape in which 
no white cell is adjacent to a black one and there is no grey cell.  This shape is formalized 
in the following definitions.  A   subset of white cells is a  \emph{white component} if it is a connected component w.r.t.
the subset of all the white cells.

\begin{definition} \label{def::good}
  A configuration
$\conf{t}$ is  \emph{regular} if   the following properties hold

\begin{description}

\item{a)} No cell is   grey.

\item{b)} Every white  component   is adjacent to a red cell.

\item{c)} No white cell   is adjacent to a black cell.
\end{description}

\end{definition}

\noindent
Observe that the starting configuration $\conf{0}$ is  regular.   

\begin{definition} \label{def::redclose}
   A white cell is  red-close  if it is adjacent to a red cell.
\end{definition}

\noindent
The next lemma 
determines the  local state-evolution of any  cell in a  regular configuration   (the proof
is  given in the Appendix).

\begin{lemma}  \label{lm::color}
    Consider any cell $\cc$ at time step $t$ and assume  $\conf{t}$  be  regular.
    Then the following properties  hold. \\
 1)   If $\cc$ is   red-close    then it  becomes red in $\conf{t+1}$ w.h.p. \\
2)  If $\cc$ is  (no red-close) white   then it    becomes white or red in $\conf{t+1}$. \\
3) If  $\cc$ is red  then it    becomes    red or black in $\conf{t+1}$. \\
4) If $\cc$ is    black  then it    becomes    red or black in $\conf{t+1}$.
\end{lemma}

\noindent
As an easy  consequence of the above lemma, we get the following

\begin{lemma} \label{cor::good}
For any $t \leqslant n$, if $\conf{t}$ is  regular, then w.h.p.  $\conf{t+1}$ is  regular as well.
\end{lemma}

\noindent
The above result on  regular configurations provides a clear characterization of the shape
of the infection process.  We now analyze the speed of the process.
Observe  that we can now assume that all configurations are  regular (w.h.p.). 
The proof of the next lemma is given in the Appendix.

\begin{lemma}\label{lm::speed}
For any $t < n$, let   $\ww$ be any white cell in $\conf{t+1}$
and let $\Red{t}$ be the set of red cells in $\conf{t}$.
 It holds w.h.p.  that
 
 \[  \cd(\ww,\Red{t+1}) \leqslant  \cd(\ww, \Red{t}) -1              \]
    
\end{lemma}
 
 \noindent
 Starting from the initial configuration $\conf{0}$ (that has one red cell), Lemma \ref{lm::speed}
 implies that \emph{every} white cell in $\sS$ w.h.p will become red within $O(\cD(\sS)) = O(D(\sS)/R)$ time steps.
 Moreover, thanks to Lemmas \ref{lm::color} and \ref{cor::good}, every red cell will become either red or black.
 Finally, when all cells are   black or red, after the next time step there are no more white agent
 and the theorem follows.

\section{ Low Transmission-Rate  or High-Mobility  I}\label{sect::high0}

\noindent
 We consider the   network formed by   $n$ mobile agents with transmission radius $R\geqslant c_0\sqrt{\log n}$   and move radius $\rho$
 such that   $R/2 \leq \rho \leq \alpha R^2/ \sqrt {\log n}$ for sufficiently small constant $\alpha>0$. The agents act on
  a support region $\sS$ that satisfies two measurable conditions. 
  
  \noindent
   It is 
$(\ell_0, \gamma)$-measurable where $\ell_0 = R/(4\sqrt 2)$ and $\gamma$ is a positive constant.
Moreover, $\sS$ is also $(\ell, \gamma)$-measurable where $\ell = 2\ell_0$ and 
 the grid of $\ell$-cells is a subgrid of the grid of cells of side length $\ell_0$. 
    We assume
  the geometric-MANET  over $\sS$ satisfies the density condition w.r.t. the $\ell_0$-cells.

 \begin{theorem} \label{thm::uppfast1}
Under the above assumptions,    the 1-flooding time
  is w.h.p. bounded by   $\Theta(D(\sS) /\rho)$.
  \end{theorem}

\noindent
The above theorem implies an optimal bound on 1-flooding time over an $L \times L$ square $\sQ$ for any sufficiently 
large $L$.

\begin{cor}\label{thm::upp square2}
Let   $R \leqslant L$ be such that $ R\geqslant c_0 L\sqrt{\log
n / n}$ for  a  sufficiently large constant   $c_0$ and let 
 $R/2 \leqslant \rho \leqslant \alpha R^2/ \sqrt{ \log n}$, for sufficiently small constant $\alpha>0$. 
  Then the  1-flooding time over $\sQ$
 is    w.h.p.   $ \Theta (L/ \rho)$.
\end{cor}

 \subsection{Proof of Theorem \ref{thm::uppfast1}} \label{ssec::proof 2}

\noindent
We   adopt      Def.s \ref{def::cellstate1},    \ref{def::good}, and    \ref{def::redclose}
given in the previous section.  Moreover, we need the further

\begin{definition}\label{def::rho-close}
Two cells are $\rho$-close if the (Euclidean) distance between  their geometric centers  is at most       $\rho$.
\end{definition}

 
 \noindent
 As in the previous section, we provide   the  local state-evolution law
  of any  cell in a  regular configuration   (the proofs of all the next 
lemmas are   given in the Appendix).

\begin{lemma}  \label{lm::color1}
    Consider any cell $\cc$ at time step $t$ and assume  $\conf{t}$  be  regular.
    Then the following properties  hold w.h.p. \\
 1)   If $\cc$ is   red-close    then it  becomes red in $\conf{t+1}$. \\
 2) If $\cc$ is $\rho$-close to a red-close cell, then $\cc$ becomes red   in $\conf{t+1}$.\\
3)  If $\cc$ is   white  but it is not $\rho$-close to any
red-close cell then it    becomes white or red in $\conf{t+1}$. \\
4) If  $\cc$ is red  or black but it is not $\rho$-close to any
red-close cell then it    becomes    red or black in $\conf{t+1}$. \\
 \end{lemma}


\noindent
As an easy  consequence of the above lemma, we get the following

\begin{lemma} \label{cor::good1}
For any $t <  n$, if $\conf{t}$ is  regular, then  w.h.p.  $\conf{t+1}$ is  regular as well.
\end{lemma}

\noindent
   In what follows, we assume that all configurations are  regular (w.h.p.) 
   and   analyze the speed of the infection process. Differently from the previous section, we will 
   show    this ``speed'' is $\Theta(\rho)$. Notice that the cell diameters of $\sS$ w.r.t. the $\ell_0$ and $\ell$ respectively  differs by a ``small''  constant factor only: for this reason both of them are denoted as $\cD(\sS)$.

\begin{lemma}\label{lm::speed1}
For any $t <  n$, let  $\ww$ be any white cell in $\conf{t+1}$ 
and let $\Redc{t}$ be the set of red-close cells in $\conf{t}$.
  It w.h.p.  holds that
 
 \[  \cd(\ww,\Redc{t+1}) \leqslant  \max\{ \cd(\ww, \Redc{t}) - \Theta(\rho / R), 0 \}              \]
    \end{lemma}

\noindent
The starting  configuration $\conf{0}$ is regular and contains one red cell. So, 
 Lemma \ref{lm::speed1}
 implies that \emph{every} white cell in $\sS$ w.h.p will become red within 
 
 \[ O\left ( \frac{ \cD(\sS) \ell}{\rho} \right ) = O\left(\frac{D(\sS)}{\rho} \right)  \ \mbox{time steps}  \]
 Moreover, thanks to Lemma \ref{lm::color1}, every red cell will become either red or black.
 Finally, when all cells are   black or red, in  the next time step there are no more white agent
 and the theorem follows.

\section{ Low Transmission-Rate  or High-Mobility  II}\label{sect::high}

 In this section, we study the case when the move radius can be much larger than the transmission  radius.
 More precisely, the move radius can be an  arbitrary polynomial of the transmission radius, i.e.
  $\rho = O(\poly(R))$,
we also assume $R \geqslant c_0 \sqrt{\log n}$, for a sufficiently large $c_0$, and $\rho \geqslant 5R$.

   \noindent
A slightly different version of the   geometric random-walk model is here adopted,  the  \emph{cellular random walk}:
 the region $\sS$ is partitioned in squared \emph{supercells} of edge length $ \rho$. Then an agent lying 
 in any position of a supercell $C$ selects her next position independently and uniformly at random over the  
 neighborhood $N(C)$. Moreover, an agent can be informed by another (red) agent
 only if they both belong to the same supercell. So,
  in the  cellular random walk model, the \emph{influence} of an agent lying in  a supercell
 $C$, in the next time step, is always restricted to the subregion $N(C)$. 
   Observe that the cellular random walk model preserves all significant 
 features  of the standard one  while, on the other hand, it avoids  several  geometric technicalities.
  The latter  would yield  a much more elaborate    analysis  without increasing the relevance of the obtained results.  
    
\noindent
We consider a support region $\sS$ that satisfies two measurable conditions. It is 
$( \ell, \gamma)$-measurable where $\ell = R/\sqrt 2$ and $\gamma$ is a positive constant.
Moreover, $\sS$ is also $(\rho, \gamma)$-measurable where the grid of supercells is a subgrid
of the grid of cells of side length $\ell$.
In what follows,      the cells of size length $\rho$ are called   \emph{supercells}    and  those of size length $\ell$
are called  cells, simply.
We also assume that the geometric-MANET over $\sS$ satisfies the density condition w.r.t. the cells.
The density condition for the stationary phase 
of the  cellular random-walk model for $\rho$ over a region  satisfying the  two above  measurability
 can be proved by similar arguments as in  \cite{CMPS09,CPS09}. Indeed, the  
  measurability  by supercells guarantees  that the stationary distribution is almost uniform; 
 this fact together with the measurability by  cells   implies the density condition.

 \begin{theorem} \label{thm::uppfast2}
Under the above assumptions and $5R \leqslant \rho \leqslant \poly(R)$,
    the 1-flooding time is w.h.p.   $\Theta(D(\sS) /\rho)$.
\end{theorem}

\noindent
The above theorem implies an optimal bound on 1-flooding time over an $L \times L$ square $\sQ$.

\begin{cor}\label{thm::upp square3}
Let   $R = o(L)$ be such that $ R\geqslant c_0 L\sqrt{\log
n / n}$ for  a  sufficiently large constant   $c_0$ and let 
 $5R \leqslant \rho \leqslant \poly(R)$. 
  Then the  1-flooding time over $\sQ$, in the cellular random-walk model with move radius $\rho$,
 is  $ \Theta (L/ \rho)$ w.h.p.
\end{cor}

 \subsection{Proof of Theorem \ref{thm::uppfast2}} \label{ssec::proof 3}

The   higher agent mobility forces us to analyze the infection process over the supercells (besides over the cells).
During the information spreading,   the state    a supercell can assume is defined by  some bounds on the number of
red and white agents inside it.  Roughly speaking, the number of white agents must  never be too small
w.r.t. the number of red agents, moreover the latter must  increase exponentially w.r.t. $R^2$.
Since the infection process is rather complex, we need to consider a relative large number of possible supercell states.
 Our analysis will first show    every supercell   eventually evolves from the initial white state to the final black
 state according to  a monotone process over a set of  intermediate states.
 Then, we will show that the speed of this process is asymptotically  optimal, i.e.,
  proportional to the move radius.

\noindent
For a supercell $C$ and a time step $t$, let   
$\nR{C}{t}$, $\nW{C}{t}$, and $\nB{C}{t}$ be, respectively, the number of red, white, and black  agents in $C$.  
 We define 
\begin{equation} \label{eq::h}
\hat h \ = \ \left\lceil   \log_{R^2 }\left( c_0 \frac{\rho^2}{R^2} \log n \right)   \right\rceil
\end{equation}
\noindent
Observe  that the assumption  $\rho = O(\poly(R))$ implies $\hat h = \Theta(1)$. 

\begin{definition}\label{def::hstate}
For any time $t$ and for any supercell $C$, we define some possible states of $C$ at time $t$.
\begin{description}
\item[State $h=0$ (White State):]   $\nR{C}{t} = 0$ \   \textsc{and} \  $\nB{C}{t} =0$.

\item[State $h = 1, \ldots , \hat h-1$ (Intermediate States):]   The values of 
 $\nR{C}{t}$ and   $\nW{C}{t}$   satisfy 
 \[ a_h R^{2h} \leqslant \nR{C}{t} \leqslant b_h R^{2h} \: \:  
\mbox{ \textsc{and} } \: \:    \nW{C}{t} \geqslant c_h \rho^2 \]
\noindent where $a_h,b_h,c_h$ are constants (they will be fixed later in the Appendix) that 
satisfy $a_h <  b_h$, $a_1 \geqslant a_2\geqslant \cdots \geqslant a_{\hat h-1} > 0$,
$0 < b_1 \leqslant b_2\leqslant \cdots \leqslant b_{\hat h-1}$, and $c_1 \geqslant c_2 \geqslant \cdots 
\geqslant c_{\hat h - 1} > 0$.

\item [State  $\hat h$ (Red State):]    $\nR{C}{t} \geqslant 90\frac{\rho^2}{R^2} \log n$. 

\item[State  $\hat h+1$ (Black State):]   $\nW{C}{t} = 0$.
\end{description}
\end{definition}

\noindent
All the above  states are mutually disjoint but the last three ones, i.e., $\hat h-1$, $\hat h$, $\hat h +1$.
For every  supercell $C$ and time step $t$,  in order to indicate that $C$ satisfies the condition of 
state $h$, we will write   $h^t(C)= h$.

\begin{definition}
The   configuration $\conf{t}$ is regular if  for every supercell $C$ (1) an $h \in \{  0,1,\ldots, \hat h, \hat h+1\}$
exists s.t.    $h^t(C) =h$
and (2)  if   $h^t(C) = \hat h +1$, it holds that,$\forall C' \in N(C)$, 
$h^t(C')  =  \hat h  \vee h^t(C')=\hat h +1$.
\end{definition}

\noindent
In the sequel, we exchange the order of the phases in  a time step: the move phase now comes before the transmission one. Clearly,
  this does not change our asymptotical results.
The next  technical lemmas allow to control the 1-step evolution of the state of any supercell in terms
of the  number of red and white agents and how such agents spread over  its cells
(the proofs are given in the Appendix).

\begin{lemma} \label{lm::wspread}
Let  $C$ be a supercell such that  $\nW{N(C)}{t} \geqslant \lambda \rho^2$, for some constant
$\lambda \geqslant 720/c_0^2$. Then, immediately after the move phase of time $t+1$ (and before
the transmission phase),  w.h.p., for every cell $\cc$ in $C$, it holds
  that the  number of white agents in   $\cc$ is at least $(\lambda/36) R^2$.
\end{lemma}
 
\begin{lemma} \label{lm::rspread}
Let  $C$ be a supercell such that  $\nR{C}{t} \geqslant \lambda R^k$, for some constants
$\lambda \geqslant 1800/c_0^2$ and $k\geqslant 2$. Then, immediately 
after the move phase of time $t+1$, w.h.p. 
 in every supercell $C' \in N(C)$, the cells in $C'$ hit by some red agent are at least 
 $\min\{(\lambda/30) R^k, \rho^2/(2R^2)\}$.
\end{lemma}

\begin{lemma}\label{lm::rspread2}
Let $C$ be any supercell such that $h^t(C) = \hat h$. Then, immediately 
after the move phase of time $t+1$, w.h.p. for every supercell $C'\in N(C)$ all the 
cells in $C'$ are hit by some red agent.
\end{lemma}

 \begin{lemma} \label{lm::uppred}
 For any time step $t$ and  any supercell $C$, let $\widehat{\nR{C}{t}} = \max\{ \nR{C'}{t}, C' \in N(C) \}$. 
 If, for some $M>0$, it holds that $\widehat{\nR{C}{t}} \leqslant M$, then it holds
  that $\nR{C}{t+1} \leqslant 68   \eta_2 M R^2$ w.h.p.
  \end{lemma}

\noindent
\noindent
We are now able to  provide   the  local state-evolution law
  of any  supercell in a  regular configuration   (the proofs of all the next 
lemmas are   given in the Appendix).
Let us define $m^t(C) = \max\{ h \  |  \exists C' \in N(C) : \  h^t(C') = h  \}$. 

\begin{lemma}\label{lm::conf}
 if  $\conf{t}$ is regular then the following implications hold w.h.p., for every supercell $C$:\\
 \begin{tabular}{llcl}
(a) &
 $m^t(C) = 0$ &$\Rightarrow$&  $h^{t+1}(C) = 0$\\
(b) &
$1 \leqslant m^t(C) \leqslant \hat h -1$  &$\Rightarrow$& $h^{t+1}(C) = m^t(C)+1$\\
(c) &
$m^t(C) = \hat h \wedge ( h^t(C) = h$ with $h < \hat h)$ & $\Rightarrow$& $h^{t+1}(C) = \hat h$\\
(d) &
$m^t(C) = \hat h \wedge h^t(C) = \hat h$ &$\Rightarrow$ & $h^{t+1}(C) = \hat h +1$\\
(e) &
$m^t(C) = \hat h +1$ &$\Rightarrow$&  $h^{t+1}(C) = \hat h+1$
 \end{tabular}
\end{lemma}
 
\noindent
As a consequence of the above lemma, we get 
\begin{lemma} \label{lm::regular}
For any $t< n$, if $\conf{t}$ is  regular, then  w.h.p.  $\conf{t+1}$ is  regular as well.
\end{lemma}

\begin{lemma}\label{lm::initial}
With high probability the initial configuration $\conf{0}$   is regular 
and a supercell $C$ exists such that $h^0(C) = 1$.
\end{lemma}

\noindent
For any time $t$, let $\mathrm{Red}(t)$ be the set of supercells whose state at time $t$ is at least 1. For any supercell
$C$, denote by $\cd^t(C)$ the distance w.r.t. supercells
 between  $C$ and   $\mathrm{Red}(t)$.  Clearly, if $C\in\mathrm{Red}(t)$
then $\cd^t(C) = 0$.

\begin{lemma}\label{lm::speed2}
For any  $t \leqslant n$,   if $\conf{t}$ is regular, then w.h.p., let  for any supercell  $W$ 
 such that $h^t(W) = 0$,
 it holds that
 \[  
\cd^{t + 1}(W) \leqslant  \cd^t(W) - 1            
 \]
\end{lemma} 

\noindent
Theorem~\ref{thm::uppfast2} is an easy consequence of Lemmas~\ref{lm::regular}, \ref{lm::initial} and \ref{lm::speed2}.

\section{The Multi-Source Case and The Completion Threshold} \label{sec::neg}

\medskip
\noindent
\textbf{The  multi-source 1-flooding.}
Consider the 1-flooding process with $n$ agents over the region $\sS$ whose starting configuration contains
an arbitrary  subset  of source agents. Every source agent has the same message (infection) and, again, 
the goal is to bound the completion time. Let $A$ be the set of positions of the source agents at starting time.

\noindent
 For any point $x\in \sS$, define  
\[ d(x,A) = \min\{ d(x,a) \ |   a \in A \} \]
\[ \ecc (A,\sS) =  \max \{  d(x,A)  \ | \ x \in \sS \}  \]
The parameter $\ecc(A,\sS)$ is the   geometric eccentricity of $A$ in $\sS$.

\begin{theorem}\label{thm::multisource}
Under the same  assumptions of the single-source case, for any choice 
of the source positions  $A$,  the 1-flooding time is w.h.p. 

\noindent
i) $\Theta(\ecc(A,\sS)/R)$,  for any $\rho \leqslant R/(2\sqrt 2)$;

\noindent
ii)  $\Theta(\ecc(A,\sS)/ \rho)$, for any $ R/2 \leqslant \rho \leqslant \poly(R)$.
\end{theorem}

\ideaproof
The crucial observation is that in all cases  (i.e. those of Sect.s \ref{ssec::slow}, \ref{sect::high0}, and \ref{sect::high}),
the obtained local state-evolution law of the cells works for \emph{any} starting configuration 
provided it is a regular one. It is easy to verify that, for any choice of the 
source subset, the starting configuration  is w.h.p. regular. Moreover,   our analysis
   of the speed of the information spreading does not change at all: the initial distance between
   every white cell (supercell) and its closest red cell (red-close supercell) decreases w.h.p. by 1 at any
   time step. In the case of more source agents, such initial distance is bounded by $\ecc(A,\sS)$.
   
 \qed

\medskip
\noindent
\textbf{The  completion threshold for $R$.}
We   show that the 1-flooding  may not complete  when  $R <  D(\sS)\sqrt{\gamma \log
n / n}$ for  a  sufficiently small constant   $\gamma$.  
By using basic probability arguments it is easy to prove that, under the above assumption,
the probability that       \emph{some} isolated agents exist in the initial configuration   is \emph{positive}.
In the next lemma we state  a much stronger result: we prove that, w.h.p., the number
of such isolated agent is $n^{\Theta(1)}$. We don't know whether 
this result has been explicitly proved somewhere, however,  in the appendix we   give a proof based
 on  the Method of Bounded Differences \cite{M89}.

\begin{lemma}\label{lm::negres}
A constant $\gamma >0$ exists such that, for sufficiently large $n$, for any  $R <  D(\sS) \sqrt{\gamma \log
n / n}$   and $\rho>0$, w.h.p. there are $ m= n^{\alpha}$ (for some constant $\alpha >0$)
 isolated agents in the initial configuration.  
\end{lemma}

 \begin{cor} \label{cor::neg}
A constant $\gamma >0$ exists such that, for sufficiently large $n$, for any  $R <   D(\sS)\sqrt{\gamma \log
n / n}$   and $\rho>0$, w.h.p., there are  source agents for which 1-flooding does not complete.

\end{cor}

\noindent
 By   using arguments similar
to those in the     proof of  Lemma \ref{lm::negres} (and a smaller $\gamma$),
    the above negative result  can be extended to the $k$-flooding 
 for any fixed constant  $k$ when $\rho \leqslant R$.

 \section{Conclusions} \label{sec::conc}
 The probabilistic analysis of Information spreading in mobile/dynamic networks is a   challenging issue that
 is the subject of several current research projects and 
some relevant advances have been obtained in the last 5 years by using approaches  based on 
time/space discrete approximation of the evolving systems and 
discrete probability \cite{BCF09,CMPS07,CMMPS08, DMP08, JMR09, PSSS10, PPPU10}.
  We believe that this  \emph{discrete} approach is   promising    to address
several important   related questions  which are still far to be solved.

\noindent
The more related  open question to our work is the analysis of  the $k$-flooding with $k>1$ under the 
   completion threshold of the 1-flooding process.
    For instance, is the $\Theta(\log n)$-flooding able to complete even  much \emph{under} the above threshold?
Which is the role of the move radius $\rho$?

\noindent
A more general challenging issue is to extend the analysis of the parsimonious flooding to other explicit models
of MANET such as the random way-point model  \cite{LV05,CMS10}.

\newpage

 \appendix
 \section{Proofs of Section \ref{ssec::slow}}

\medskip
 \noindent
\textbf{Proof of Lemma \ref{lm::color}.}  

\noindent
1) After the transmission phase of time step $t+1$, every agent in the red-close cell $\cc$ becomes red.
Then, thanks to the density condition, w.h.p. at least a fraction of these red agents remains
in $\cc$ after the move phase as well. Thus, at time $t+1$, w.h.p. $\cc$ becomes red.

 \medskip
 \noindent
2)
Observe that if a white cell $\cw$ is not red-close then, since $\conf{t}$ is  regular,  $\cw$
is surrounded by white cells only, then at time $t+1$ no black agent can reach $\cw$.
Notice that $\cw$ may become red at time $t+1$. Indeed, if there is a red-close cell in $N(\cw)$, after the move phase
of time step $t+1$, 
 some  red agent can reach $\cw$.

 \medskip
   \noindent
3)
 After the transmission phase of step $t+1$, since $R = 2\sqrt 2 \ell$, all agents in $N(\cc)$ will be either red or black.
 So, after the move phase, since $\rho < \ell$ no white agent can reach    $\cc$.

 \medskip
  \noindent
4) Since $\conf{t}$ is  regular,  a black cell $\bb$ is initially  surrounded by black or red cells only.
   After the transmission phase of step $t+1$, all agents in $N(\bb)$ will be either red or black.
 So, after the move phase, since $\rho < \ell$ no white agent can reach    $\bb$.
 
 \qed
 
\medskip
  \noindent
\textbf{Proof of Lemma \ref{lm::speed}.} 
   From Lemma \ref{lm::color},  a white cell $\ww$ in   $\conf{t+1}$ 
   was white   in $\conf{t}$ as well.  Now, consider one   shortest cell path $p$ from $\ww$ to a red cell $\rr$ in  $\Red{t}$.
    Then, thanks to Lemma \ref{lm::color},
   the
    (red-close) white cell $\ww'$ in $p$ and adjacent to $\rr$  will w.h.p. become red in $\conf{t+1}$.
     Notice that some other white cell in $p$ may
    become red in $\conf{t+1}$. In any case, the   lemma  holds.
    
 \qed


\section{Proofs of Section \ref{sect::high0}}

For the sake of simplicity, we here assume   the parameter  $\gamma$ of  the measurability of region $\sS$ be
set to 1.  
If $\gamma <1$,  only the values of the constants in the proofs would change.

\noindent
\textbf{Proof of Lemma \ref{lm::color}.} 

\noindent \emph{Claim 1).}  Immediatly after the transmission phase of time step $t$, all agents in $\cc$ become red.
Thanks to the density condition and the bound $\rho \leq \alpha (R^2/\sqrt{\log n})$, it is possible to prove that
w.h.p at least one of such red agents remains in $\cc$ after the move phase.

\noindent \emph{Claim 2). }
We first observe that, after the transmission phase of $t+1$, all   agents in $\cc'$ becomes red. 
Define $X$ be the indicator r.v. that equals 1 iff $\cc$ is hit by some red agent immediatly after
the move phase.
From the density condition, it holds that 

\[  \Prob{X=0}  \leqslant \left(1 - \frac{\ell_0^2}{\pi{\rho}^2}\right)^{\eta_1\ell_0^2} \leqslant \exp\left(- \frac{\eta_1\ell_0^4}{\pi \rho^2}\right) \]
The upper bound   $\rho \leqslant (R^2/\sqrt{\log n})$ implies that

\[  \Prob{X=0}  \leqslant \exp(- \Theta(\log n)) \]
So, w.h.p.  cell $\cc$ becomes red.

\noindent \emph{Claim 3). } Assume, by way of contradiction, that immediatly after   the transmission
phase of $t+1$, a black agent   exists that is within Euclidean distance $\rho$ from $\cc$.
Let $b$ the black agent closest to $\cc$ and 
let $\bb$ containing it. Let $\cz$ be the cell which is the closest one  to $\cc$ among those in   $N(\bb)$. 
Since the region is convex, by simple geometric argument, we have that 
the Euclideand distance between the centers of $\cc$ and $\cz$ is equal to the Euclidean distance
between $\bb$ and $\cc$. This implies that $\cc$ is $\rho$-close to $\cz$.
Since  $\conf{t}$ is regular, it is not hard to prove that cell $\cz$ can be neither black nor white
at time step $t$. It turns out that $\cz$ must be red: this would contradict the hypothesis that
$\cc$ is not $\rho$-close to a red-close cell at time step $t$.

\noindent \emph{Claim 4). } 
Immediatly after   the transmission
phase of $t+1$, consider the  white  agent   $w$ closest to cell  $\cc$.
If $d(w,\cc)>0$, we easily get the claim. Assume thus $d(w,\cc) \leqslant \rho$ and let
$\cw$ be the cell containing $w$. Let    $\cz$ be 
 the cell which is the closest one  to $\cc$ among those in   $N(\cw)$. 
 Since    $\conf{t}$ is regular, it is not hard to prove that cell $\cz$ can be neither red nor black
at time step $t$. So, cell $\cz$ must be white at time $t$. After the transmission of time step
$t+1$, all agents in $\cz$ must go into the red state  since otherwise $w$ would not be the closest white
agent to $\cc$.  Now, since $\cz$ is $\rho$-close to $c$, by using the same argument of Claim 2,
we have that after the move phase,  at least a red agent hits cell $\cc$ thus the latter becomes red as well.

 \qed

\medskip
\noindent
\textbf{Proof of Lemma \ref{lm::speed1}.}
From Lemma \ref{lm::color1},  a white cell $\ww$ in   $\conf{t+1}$ 
   was white   in $\conf{t}$ as well.  
  Consider one   shortest (white) cell path $p(\ww,\rr)$ from $\ww$ to $\Redc{t}$    in $\conf{t}$.
    Then, thanks to Claim 2 of Lemma \ref{lm::color1},
every      
        cell $\ww'$ in $p(\ww,\rr)$ which is  $\rho$-close      to $\rr$  will w.h.p. become red in $\conf{t+1}$.
        Notice that the number of such new red cells is at least $\rfloor \rho/(\sqrt 2  \ell)$.
     Notice that some other white cell in $p(\ww,\rr)$ may
    become red in $\conf{t+1}$. In any case, the   lemma holds.

\qed

\section{Proofs of Section~\ref{sect::high}}
 
For the sake of simplicity, we here assume   the parameter  $\gamma$ of  the measurability of region $\sS$ be
set to 1.  
If $\gamma <1$,  only the values of the constants in the proofs would change.

\medskip

 \noindent
\textbf{Proof of Lemma \ref{lm::wspread}.}
 Let $X$ be the number of white agents in cell $\cc$ immediatly after  the move phase of time step $t+1$.
 At time $t$, for every white agent $j$ in $N(C)$, let $Y_j$ be the indicator r.v. that is 1 iff agent $j$ moves to 
 cell $\cc$. Notice that  $\Prob{Y_j = 1} \geq R^2/(18 \rho^2)$.
 Clearly,  $ X = \sum_j Y_j$ and 
 
 \[ \Expec{X} \geqslant    \nW{N(C)}{t} \frac{R^2}{18 \rho^2}  \geqslant  \frac{\lambda}{18} R^2 \]
 
 Since the r.v. $Y_j$ are independent, we can apply Chernoff's Bound and get that, w.h.p.
 $X \geq  \frac{\lambda}{36} R^2$. The lemma follows by applying the union bound with respect to all cells in $C$.
 \qed

\medskip

\noindent
\textbf{Proof of Lemma \ref{lm::rspread}.}
We distinguish two cases. Firstly consider the case that 
\begin{equation}\label{ineq::rhobound}
\rho \geqslant c_0R\sqrt{\frac{\lambda}{18}\log n}
\end{equation}
Let $E$ be a subset of the red agents in $C$ at time $t$ of cardinality $M = \min\{ \lambda R^k, (18\rho^2)/R^2\}$.
Let $Z$ be the number of cells in $C'$ hit by some red agent in $E$ immediately after the move phase at time $t + 1$. We first bound
 the expected value of $Z$.  For any cell $\cc$ in $C'$, let $Y_{\cc}$ the indicator r.v. that is 1 iff $\cc$ is hit.
 It holds that
 \[ 
 \Prob{Y_{\cc} =0}  \leqslant \left(  1-\frac{R^2}{18\rho^2}\right)^M  \leqslant \exp\left(-\frac{
 R^2}{18\rho^2}M\right)\]
Hence, 
 \[ \Prob{Y_{\cc} =1} \geqslant  1 - \exp\left(-\frac{R^2}{18\rho^2}M\right)      \geqslant  
(1 - 1/e)\frac{R^2}{18\rho^2}M 
\]
 where the second inequality derives from the inequality $\exp(-x) \leqslant 1 - (1 - 1/e)x$, for any $0 \leqslant x \leqslant 1$.
 We then get 
 \begin{equation*} 
\Expec{Z}  = \sum_{\cc}\Expec{Y_{\cc}} \geqslant \frac{2\rho^2}{R^2} (1 - 1/e)\frac{R^2}{18\rho^2}M 
\geqslant \frac{M}{15}
\end{equation*}
\noindent
Notice that r.v. $Y_{\cc}$ are not independent. 
So, in order to prove a concentration result for $Z$, we use the \emph{method of bounded differences}
\cite{M89}.
For every red agent $j$ in $E$, let $X_j$ be the r.v. that is equal to the cell $\cc$ in $C'$ hit by
agent $j$ immediately after the move phase of time $t+1$ and $X_j = 0$ if $j$ does not hit any cell in $C'$. Observe that   r.v. $X_j$ are independent.  Now, consider function  $F(X_1,\ldots,X_M)$ that returns
the number of cells in $C'$ hit by some red agent. It is easy to see that $F$ has   Lipschitz constant
equal to 1.  Moreover it holds that $F(X_1,\ldots,X_M) = Z$. By applying Lemma 1.2 in \cite{M89}, we get 
\begin{eqnarray*}
\Prob{Z \leqslant \min\left\{\frac{\lambda R^k}{30}, \frac{\rho^2}{2R^2}\right\}} & \leqslant &
\Prob{Z \leqslant \frac{M}{30}} \leqslant \Prob{Z \leqslant \frac{1}{2}\Expec{Z}}\\
&  \leqslant & 2\exp\left(-\frac{M}{450}\right)
\end{eqnarray*}
From Inequality~(\ref{ineq::rhobound}), the assumption $R \geqslant c_0 \sqrt{\log n}$, and the lemma's hypothesis
 on $\lambda$,
it is easy to see that we get high probability.
The thesis follows by applying the union bound with respect to all  the supercells in $N(C)$.

\noindent Now consider the case that
\begin{equation}\label{ineq::rhobound2}
\rho < c_0R\sqrt{\frac{\lambda}{18}\log n}
\end{equation}
Define 
\[
q = \left\lfloor \sqrt{\frac{\lambda}{18}}\frac{R^2}{\rho} \right\rfloor
\]
Notice that $q \geqslant 1$, since Inequality~(\ref{ineq::rhobound2}) holds.
We partition every cell into subcells of side length $s = \ell/q$. Let $m = (9\rho^2)/s^2$. Observe that $m$ is an upper bound 
on the number of subcells in $N(C)$. It holds that
\[
 m = \frac{9\rho^2}{s^2} = \frac{9\rho^2}{\ell^2}q^2 \leqslant \frac{18\rho^2}{R^2}\left(\sqrt{\frac{\lambda}{18}}\frac{R^2}{\rho}
 \right)^2 = \lambda R^2
\]
Thus, $m \leqslant \lambda R^k$. Let $E$ be a subset of the red agents in $C$ at time $t$ of cardinality $m$.
Let $Z$ be the number of subcells in $C'$ hit by some red agent in $E$ immediately after the move phase at time $t + 1$. We first bound
 the expected value of $Z$.  For any subcell $\mbox{\sc{s}}$ in $C'$, let $Y_{\mbox{\sc{s}}}$ the indicator r.v. that is 1 
 iff $\mbox{\sc{s}}$ is hit.
 It holds that
 \[ 
 \Prob{Y_{\mbox{\sc{s}}} =0}  \leqslant \left(  1-\frac{s^2}{9\rho^2}\right)^m  \leqslant \exp\left(-\frac{
 s^2}{9\rho^2}m\right) = \frac{1}{e}\]
Hence, $\Prob{Y_{\mbox{\sc{s}}} =1} \geqslant  1 - 1/e$.
 We then get 
 \begin{equation*} 
\Expec{Z}  = \sum_{\mbox{\sc{s}}}\Expec{Y_{\mbox{\sc{s}}}} \geqslant \frac{\rho^2}{s^2} (1 - 1/e)
\geqslant \frac{m}{15}
\end{equation*}
\noindent
For every red agent $j$ in $E$, let $X_j$ be the r.v. that is equal to the subcell $\mbox{\sc{s}}$ in $C'$ hit by
agent $j$ immediately after the move phase of time $t+1$ and $X_j = 0$ if $j$ does not hit any subcell in $C'$. 
Observe that   r.v. $X_j$ are independent.  Now, consider function  $F(X_1,\ldots,X_m)$ that returns
the number of subcells in $C'$ hit by some red agent. Clearly $F$ has   Lipschitz constant
equal to 1 and $F(X_1,\ldots,X_m) = Z$. By applying Lemma 1.2 in \cite{M89}, we get 
\[
\Prob{Z \leqslant \frac{m}{30}} \leqslant \Prob{Z \leqslant \frac{1}{2}\Expec{Z}}
 \leqslant 2\exp\left(-\frac{m}{450}\right)
\]
It is not hard to see that $m \geqslant (\lambda R^2)/4$. From this and the assumed bound on $\lambda$, we get the high
probability. Observe that the number of cells in $C'$ hit by red agents is at least $Z/q^2$. This implies that, w.h.p., the number
of cells in $C'$ hit by red agents is at least 
\[
\frac{m}{30}\frac{1}{q^2} \geqslant \frac{9\rho^2}{30s^2q^2} = \frac{9\rho^2}{30\ell^2} = \frac{18\rho^2}{30R^2} \geqslant 
\frac{\rho^2}{2R^2} \geqslant  \min\left\{\frac{\lambda R^k}{30}, \frac{\rho^2}{2R^2}\right\}
\]
The thesis follows by applying the union bound with respect to all  the supercells in $N(C)$.
 \qed

 \medskip

\noindent
\textbf{Proof of Lemma \ref{lm::rspread2}.}
By hypothesis $\nR{C}{t} \geqslant K$ where
\[
K = 90 \frac{\rho^2}{R^2}\log n
\]
Let $\cc$ be any cell in $N(C)$ and let $Y$ be the indicator r.v. that is 1 iff cell $\cc$ is hit by some red agent
immediately after the move phase of time $t + 1$. It holds that
\[
 \Prob{Y =0} \leqslant \left(1 - \frac{\ell^2}{9\rho^2}\right)^K \leqslant \exp\left(-\frac{\ell^2}{9\rho^2}K\right)
 = \exp\left( - \frac{R^2}{18\rho^2}90\frac{\rho^2}{R^2}\log n\right) = e^{-5\log n}
\]
The thesis follows by applying the union bound with respect to all the cells in $N(C)$.
\qed

 \medskip
\noindent
\textbf{Proof of Lemma \ref{lm::uppred}.}
Assume that $\widehat{\nR{C}{t}}  \leqslant M$, then 
$\nR{N(C)}{t} \leqslant 9 M$. So,  immediately after  the move phase at time $t+1$, the
red agents in $C$ are at most $9 M$. Notice that   any red agent can 
transmit over  at most 15 cells. From the density condition, w.h.p. every cell contains at most $\eta_2\ell^2$
agents. It thus follows that w.h.p. 

\[  \nR{C}{t+1}  \leqslant  9M \cdot 15 \eta_2 \ell^2 \leqslant  68   \eta_2 M R^2
\]
\qed

\medskip
 \noindent
 \textbf{Setting the constants in Def. \ref{def::hstate}.}
 We can now  set constants $a_h$, $b_h$, and $c_h$ for $h = 1,2,\ldots, \hat h - 1$:
 \begin{eqnarray}
 \label{def::a}
 a_h & = & \frac{\eta_1^h}{2\cdot 2160^{h - 1} 20^{\frac{(h - 1)(h - 2)}{2}}}\\
 \label{def::b}
 b_h & = & 15\cdot 68^{h - 1}\eta_2^h\\
 \label{def::c}
 c_h & = & \frac{\eta_1}{2\cdot 20^{h - 1}}
 \end{eqnarray}
 
 \medskip
 
\noindent
\textbf{Proof of Lemma \ref{lm::conf}.}

\noindent
\emph{Implication (a).}\quad 
It directly follows from the definitions of state $0$ and of cellular random-walk model.

\smallskip

\noindent \emph{Implication (b).}\quad Let $m = m^t(C)$ and let $\bar C \in N(C)$ be a supercell
such that $h^t(\bar C) = m$. Since $1 \leqslant m \leqslant \hat h - 1$, from the definition of intermediate states
it derives that $\nW{N(C)}{t} \geqslant c_m \rho^2$. From Lemma~\ref{lm::wspread} (it holds that $c_m \geqslant 720/c_0^2$, 
for sufficiently large $c_0$)
it follows that, immediately
after the move phase of time $t + 1$, every cell in $C$ contains at least $(c_m/36)R^2$ white agents, w.h.p.
 
\noindent
 Since $h^t(\bar C) = m$, it holds that $\nR{\bar C}{t} \geqslant a_mR^{2m}$. 
In virtue of Lemma~\ref{lm::rspread} applied to $\bar C$  (it holds that $a_m \geqslant 1800/c_0^2$, for sufficiently large $c_0$) 
 we obtain that, immediately after the move phase of time $t + 1$,
w.h.p. the cells in $C$ hit by red agents are at least $\min\{(a_m/30)R^{2m}, \rho^2/(2R^2)\}$. 
To summarize, w.h.p., immediately before the transmission phase of time $t + 1$, every cell 
in $C$ contains at least $(c_m/36)R^2$ white agents and at least $\min\{(a_m/30)R^{2m}, \rho^2/(2R^2)\}$ of those cells are hit by
red agents. After the transmission, all the white agents in any hit cell get red, thus, w.h.p., it holds that 
\begin{equation}\label{ineq::rlb0}
\nR{C}{t + 1} \geqslant \frac{c_m}{72}\min\left\{\frac{a_m}{15}R^{2m + 2}, \rho^2\right\}
\end{equation}
 \noindent
We distinguish two cases. Firstly consider the case that 
$m \leqslant \hat h - 2$.
Notice that from the definition of $\hat h$ it holds that $(a_m/15)R^{2m + 2} \leqslant (a_m/15) R^{2\hat h - 2} 
\leqslant \rho^2$. Thus inequality (\ref{ineq::rlb0}) becomes
\[
\nR{C}{t + 1} \geqslant \frac{c_m a_m}{1080}R^{2m + 2}
\]
From the definitions~(\ref{def::a}) and (\ref{def::c}) it holds that $a_{h + 1} \leqslant a_h c_h /1080$ (for $h = 1,\ldots \hat h - 2$). Thus, the lower bound on red agents of state $m + 1$ is satisfied:
\begin{equation}\label{ineq::rlb}
\nR{C}{t + 1} \geqslant  a_{m + 1} R^{2(m + 1)}
\end{equation}
 Now we prove the upper bound. Since $\conf{t}$ is regular and for any $C'\in N(C)$ $h^{t}(C') \leqslant m$,
it holds that
\[ \nR{C'}{t} \leqslant b_m R^{2m} \]
From Lemma~\ref{lm::uppred}, since $\widehat{\nR{C}{t}} \leqslant  b_m R^{2m}$, it derives that, w.h.p. 
\[
\nR{C}{t + 1} \leqslant 68\eta_2 b_m R^{2m + 2}
\]
From definition~(\ref{def::b}), it holds that $b_{h + 1} \geqslant 68\eta_2 b_h$ (for $h = 1,\ldots \hat h - 2$). Thus, the upper bound on red agents of
state $m + 1$ is satisfied:
\begin{equation}\label{ineq::rub}
\nR{C}{t + 1} \leqslant  b_{m + 1} R^{2(m + 1)}
\end{equation}
It remains to show the lower bound on white agents. Let $W$ be the number of white agents in the supercell $C$ 
immediately after the move phase of time $t + 1$. The white agents in $C$ at time $t + 1$ are those that were in $C$
immediately after the move phase and that are not got red (after the transmission phase). It follows that
\[
\nW{C}{t + 1} \geqslant W - \nR{C}{t + 1}
\]
Recall that immediately
after the move phase of time $t + 1$, every cell in $C$ contains at least $(c_m/36)R^2$ white agents, w.h.p.
Thus, w.h.p., it holds that
\[
W \geqslant \frac{2\rho^2}{R^2}\frac{c_m}{36}R^2 = \frac{c_m}{18}\rho^2
\]
By combining this inequality with inequality (\ref{ineq::rub}), we have that (w.h.p.)
\[
\nW{C}{t + 1} \geqslant \frac{c_m}{18}\rho^2 - b_{m + 1} R^{2(m + 1)}
\]
From the definition of $\hat h$ and the assumption that $m \leqslant \hat h - 2$, it holds that
\[
R^{2(m + 1)} \leqslant R^{2\hat h - 2} \leqslant c_0 \frac{\rho^2}{R^2} \log n \leqslant \frac{\rho^2}{c_0}
\]
Hence, 
\[
\nW{C}{t + 1} \geqslant \frac{c_m}{18}\rho^2 - b_{m + 1}\frac{ \rho^2}{c_0} = 
\left(\frac{c_m}{18} - \frac{b_{m + 1}}{c_0}\right)\rho^2
\]
From definitions (\ref{def::b}) and (\ref{def::c}), it holds that $c_{h + 1} \leqslant c_h/18 - b_{h + 1}/c_0$ 
(for $h = 1,\ldots \hat h - 2$), for sufficiently large $c_0$. Thus, the lower bound on white agents of state $m + 1$ is satisfied:
\begin{equation}\label{ineq::wlb}
\nW{C}{t + 1} \geqslant  c_{m + 1} \rho^2
\end{equation}
Finally, from inequalities (\ref{ineq::rlb}), (\ref{ineq::rub}), and (\ref{ineq::wlb}), it derives that, w.h.p., $h^{t+1}(C) = m + 1$.

\noindent We now consider the case $m = \hat h - 1$. From inequality (\ref{ineq::rlb0}) it holds that (w.h.p.)
\begin{eqnarray*}
\nR{C}{t + 1} & \geqslant & \frac{c_{\hat h - 1}}{72}\min\left\{\frac{a_{\hat h - 1}}{15}R^{2\hat h}, \rho^2\right\}\\
& \geqslant & \frac{c_{\hat h - 1}}{72}\min\left\{\frac{a_{\hat h - 1}}{15}c_0 \frac{\rho^2}{R^2} 
\log n, \rho^2\right\}\\
& = & c_0 \frac{c_{\hat h - 1}a_{\hat h - 1}}{15\cdot 72}\frac{\rho^2}{R^2} \log n \qquad\quad \mbox{(since 
$\frac{a_{\hat h - 1} c_0\log n}{15R^2} \leqslant \frac{a_{\hat h - 1}}{15c_0} \leqslant 1$)}\\
& \geqslant & 90 \frac{\rho^2}{R^2} \log n\qquad\quad \mbox{(for sufficiently large $c_0$)}
\end{eqnarray*}
and thus $h^{t+1}(C) = \hat h = m + 1$.

\smallskip

\noindent \emph{Implication (c).}\quad
Since $m^t(C) = \hat h$, there is a supercell $\bar C$ adjacent to $C$ such that $h^t(\bar C) = \hat h$.
From Lemma~\ref{lm::rspread2}, immediately after the move phase of time $t + 1$, w.h.p. all the cells in $C$ are 
hit by some red agent. Since $h^t(C) < \hat h$ and $\conf{t}$ is regular, it holds that $\nW{C}{t} \geqslant c_{\hat h - 1}\rho^2$.
It follows that immediately after the transmission phase, w.h.p. the number of red agents in $C$ is at least
\[
c_{\hat h - 1}\rho^2 \geqslant 90\frac{\rho^2}{R^2}\log n
\]
for sufficiently large $c_0$. This implies that $h^{t + 1}(C) = \hat h$.

\smallskip

\noindent \emph{Implication (d).}\quad
Since $h^t(C) = \hat h$, from Lemma~\ref{lm::rspread2}, immediately after the move phase of time $t + 1$, w.h.p. 
all the cells in $C$ are hit by some red agent. It follows that immediately after the transmission phase there are no
white agents in $C$. Thus, w.h.p. $\nW{C}{t + 1} = 0$ and $h^{t + 1}(C) = \hat h + 1$.

\smallskip

\noindent \emph{Implication (e).}\quad
Let $\bar C\in N(C)$ be a supercell such that $h^t(\bar C) = \hat h + 1$. Since $\conf{t}$ is regular, it must be the case
that $h^t(C) \geqslant \hat h$ (given that $C \in N(\bar C)$). This implies that either $h^t(C) = \hat h$ or $h^t(C) = \hat h + 1$.
We distinguish the two cases.
Firstly, if $h^t(C) = \hat h$, from Lemma~\ref{lm::rspread2}, immediately after the move phase of time $t + 1$,
w.h.p. all the cells in $C$ are hit by some red agent. It follows that immediately after the transmission phase there are
no white agents in $C$. Thus $h^{t + 1}(C) = \hat h + 1$.

\noindent
Consider now the case $h^t(C) = \hat h + 1$. Since $\conf{t}$ is regular, $\forall C'\in N(C)$ we have 
 $h^t(C') =  \hat h \vee h^t(C') =  \hat h +1 $.
We distinguish two cases. 
\begin{description}
\item[1) $\exists C'\in N(C) : h^t(C') = \hat h$:] 
From Lemma~\ref{lm::rspread2}, it follows that $h^{t+ 1}(C) = \hat h + 1$.
\item[2) $\forall C'\in N(C)$ $ h^t(C') = \hat h + 1$:]
It holds that $\nW{N(C)}{t} = 0$ and thus no white agent can move into supercell $C$ during time step $t + 1$. It follows 
that $h^{t+ 1}(C) = \hat h + 1$.
\end{description}
\qed

 \medskip
 
\noindent
\textbf{Proof of Lemma \ref{lm::regular}.}
From Lemma~\ref{lm::conf} it is easy to verify that if $\conf{t}$ is regular then, w.h.p. for every supercell $C$,
it holds that $h^{t + 1}(C) \in \{0,1,\ldots, \hat h, \hat h + 1\}$.

\noindent
Let $C$ be a supercell such that $h^{t + 1}(C) = \hat h + 1$. From Lemma~\ref{lm::conf} it must be the case 
that either $h^t(C) = \hat h$ or $h^t(C) = \hat h + 1$. Distinguish the two cases.
\begin{description}
\item[1) $h^t(C) = \hat h$:] 
It holds that $\forall C'\in N(C)\; m^t(C') \geqslant \hat h$. From Lemma~\ref{lm::conf}
(c), (d), and (e), it follows that   $h^{t+1}(C') =  \hat h \vee h^{t+1}(C') =  \hat h +1 $
\item[2) $h^t(C) = \hat h + 1$:]
It holds that $\forall C'\in N(C)\; m^t(C') \geqslant \hat h + 1$. From Lemma~\ref{lm::conf} (e) it follows that
$h^{t + 1}(C') = \hat h + 1$.
\end{description}
\qed

 \medskip
 
\noindent
\textbf{Proof of Lemma \ref{lm::initial}.}
At the very beginning of time step $0$, all the agents are white except the source one which is red.
Immediately after the move phase the source agent moves to a cell $\cc$.
Immediately after the transmission phase, w.h.p. all the agents  (except one) in the cell $\cc$ get red.
Let $C$ be the supercell containing cell $\cc$. From the density condition, it holds that (w.h.p.)
\[
\eta_1\ell^2 \leqslant \nR{C}{0} \leqslant 15\eta_2\ell^2
\]
since at most 15 cells are affected by the transmission of the source agent. Thus, the bounds on red agents 
of state 1 are satisfied: $a_1R^2 \leqslant  \nR{C}{0} \leqslant b_1R^2$, since $a_1 = \eta_1/2$ and $b_1 = 15\eta_2$.

Consider now the number of white agents. In the supercell $C$ there are at least $\rho^2/\ell^2 - 15$ cells that are 
not affected by the transmission. From the density condition, those cells contain at least $\eta_1\ell^2(\rho^2/\ell^2 - 15)$
white agents. It follows that, w.h.p.
\[
\nW{C}{0} \geqslant \eta_1\ell^2(\frac{\rho^2}{\ell^2} - 15) = \eta_1\rho^2 - 15\eta_1\ell^2 \geqslant \frac{\eta_1}{2}\rho^2
\]
where the last inequality holds because $\rho \geqslant 5R$. Hence $\nW{C}{0} \geqslant c_1 \rho^2$ and $h^0(C) = 1$.
Clearly all the other supercells are in the state $0$.
\qed

 \medskip
 
\noindent
\textbf{Proof of Lemma \ref{lm::speed2}.}
Let $P$ be a shortest path of supercells between $W$ and $\mathrm{Red}(t)$. 
Path $P$ contains 
a supercell $W'$ such that $h^t(W') = 0$ and $W'$ is adjacent to a supercell in $\mathrm{Red}(t)$.
From Lemma~\ref{lm::conf} it follows that  the state of $W'$ at time $t+1$ is at least 1
 and thus $W'\in \mathrm{Red}(t + 1)$. 
Hence, $\cd^{t + 1}(W) \leqslant \cd^{t}(W)$.
\qed


\section{Proof of Section \ref{sec::neg}}

\textbf{Proof of Lemma \ref{lm::negres}.}
For the sake of simplicity, we consider $\sS$ as the square of edge length $\sqrt n$ and set 
$R \leqslant \sqrt{\gamma   \log n}$ where $\gamma$ is a suitable positive constant.
At time $t=0$, consider the random disk graph $G(n,R)$ formed by the $n$ agents.
Define the indicator r.v.  $X_i$ which equals 1 iff   agent $i$ is isolated in $G(n,R)$.
We have that

\[  \Prob{X_i = 1} \geqslant \left(1-\frac{\pi R^2}{n}\right)^{n-1} \geqslant \ \exp(-2\pi \gamma \log n) \]
This implies that

\[  \Expec{X = \sum_i X_i} \ \geqslant n^{1-2 \pi \gamma} \]
Unfortunately,  r.v. $X_i$'s are not independent so we cannot apply Chernoff's bound:
we instead apply the Method of Bounded Differences \cite{M89}.
Consider the r.v.  $Z_i$ that returns the position (i.e. coordinates) of agent $i$ in $\sS$. Observe that 
such r.v. are independent.
Then we define the function $F(Z_1,Z_2, \ldots, Z_n)$ that returns the number of isolated agents.
Observe that  $ Z =  X$.
We now need the following useful Lemma 1.2 in   \cite{M89}.

\noindent
In order to apply this    lemma to our function $F$, we need to find out the correct values for coefficients
Lipschitz's constants $c_k$. 
 We claim that, in our geometric  case, $c_k \leqslant 7$: indeed,   changing     the position of any agent in the
square can affect   the $R$-disks of at most $6$ isolated nodes.
So, by setting this bound on the $c_k$ and $t = \beta \Expec{F(Z_1,\ldots, Z_k, \ldots, Z_n)}$ for a suitable small constant
$\beta>0$,  we can choose $\gamma>0$ such that 

\[  \Prob{  X \leqslant n^{1/2}}  = \Prob{F(Z_1,\ldots, Z_n) \leqslant n^{1/2}}
 \leqslant \exp\left(  - n^{\Theta(1)}\right) \]

\qed

\end{document}